\documentclass[aps,11pt,nofootinbib]{revtex4-1}
\pdfoutput=1
\usepackage[utf8]{inputenc}
\usepackage{graphicx}
\usepackage[margin=1in]{geometry}
\usepackage{amsmath}
\usepackage{amssymb}
\usepackage{float}
\usepackage[all]{nowidow}
\usepackage[linktocpage=true,colorlinks=true,linkcolor=blue,citecolor=blue,urlcolor=blue]{hyperref}
\hypersetup{pdftitle={An accurate model for the primordial black hole mass distribution from a peak in the power spectrum},pdfauthor={3 authors}}

\newcommand{\PBH}{\mathrm{PBH}}
\newcommand{\diffd}{\mathrm{d}} 
\newcommand{\ddv}[2]{\frac{\diffd #1}{\diffd #2}} 
\newcommand{\Msun}{\mathrm{M}_\odot}

\begin{document}

\author{Andrew D.~Gow$^1$}
\email{A.D.Gow@sussex.ac.uk}
\author{Christian T.~Byrnes$^1$}
\email{C.Byrnes@sussex.ac.uk}
\author{Alex Hall$^2$}
\email{ahall@roe.ac.uk}

\affiliation{\\1) Department of Physics and Astronomy, University of Sussex,\\Brighton BN1 9QH, United Kingdom\\}

\affiliation{\\2) \mbox{Institute for Astronomy, University of Edinburgh,} Royal Observatory, Blackford Hill,\\Edinburgh, EH9 3HJ, United Kingdom\\}

\date{26/10/2021}

\title{An accurate model for the primordial black hole mass distribution from a peak in the power spectrum}

\begin{abstract}
We examine the shape of the primordial black hole mass distribution arising from a peak in the primordial power spectrum. In light of improvements to the modelling, we revisit the claim that the effects of critical collapse produce a distribution that is not described by the commonly assumed lognormal, showing that this conclusion remains valid, particularly for narrow peaks where the shape of the mass distribution is insensitive to the peak properties and critical collapse determines a minimum width. We propose some alternative models that may better describe the shape, both for the narrow peak case and for much broader peaks where the effect of the peak shape is significant. We highlight the skew-lognormal and a generalised model motivated by the physics of critical collapse as the best of these possible alternatives. These models can be used as an accurate and fast approximation to the numerically calculated mass distribution, allowing for efficient implementation in an MCMC analysis. We advocate the use of one of these two models instead of the lognormal with sufficiently accurate data, such as future LIGO--Virgo observations, or when considering strongly mass dependent constraints on the PBH abundance.
\end{abstract}

\maketitle

\section{Introduction}

Since the idea of primordial black holes (PBHs) was first postulated half a century ago \cite{Zel'dovich_1967,Hawking_1971,Carr_1974}, a lot of progress has been made in studying constraints on their abundance as well as possible signs that they have been detected. Until quite recently, most constraints on the PBH abundance assumed a monochromatic mass distribution which has the advantage of simplicity, since this is the unique case where a constraint at any given mass can be made without considering the constraints on other, similar masses. See e.g.~\cite{Carr:2009jm,Green_2015,Carr:2016drx,Sasaki_2018,Carr:2020gox} for reviews. However, the phenomenon of critical collapse means that a range of PBH masses are generated from large amplitude perturbations re-entering the horizon even if the perturbation spectrum has power at only one wavenumber \cite{Niemeyer:1998_CC,Yokoyama:1998xd,Musco:2009_CC,Kuhnel:2015vtw}, due to the spread in amplitudes of modes at that scale. Therefore, as one would intuitively expect, a monochromatic mass distribution is not physically realistic, no matter how narrowly peaked the primordial power spectrum might be.\footnote{In practise there is also a limit to how narrow the primordial power spectrum can be, with the limits depending on the model of inflation, see e.g.~\cite{Cai:2018tuh,Byrnes:2018txb,Carrilho:2019oqg,Ozsoy:2019lyy,Ashoorioon:2019xqc,Palma:2020ejf,Fumagalli:2020adf}.}

While the community was focused on making order-of-magnitude constraints to the PBH abundance and simple ``yes/no'' answers to whether PBHs of a given mass could constitute the entirety of dark matter, the approximation of a monochromatic mass distribution was adequate. However, in recent times there has been a vigorous debate about exactly what fraction of the dark matter could be contained in PBHs with a mass of order the solar mass, for example to fit to lensing surveys or the LIGO--Virgo detection of gravitational waves. Many, but not all, constraints allowed an order one fraction of PBHs to be in dark matter. See e.g.~\cite{Carr:2020xqk,Green:2020jor} for very recent reviews. These constraints come from a wide range of methods as well as probing a wide range of redshifts, and there is the possibility that accretion makes the constraints time dependent in a mass dependent manner \cite{Bosch-Ramon:2020pcz,DeLuca:2020fpg,DeLuca_2020}. Finally there are some hints that LIGO--Virgo may have detected PBHs, for example due to the low spin of most of the detected events \cite{Fernandez:2019kyb,DeLuca:2020bjf} as well as some objects which fall into or close to the lower and upper mass gaps commonly considered for astrophysical formation channels \cite{Abbott:2020uma,Abbott:2020khf,Vattis:2020iuz,LIGO:2020_190521,Clesse:2020ghq}, although these can be explained with specific astrophysical models \cite{Belczynski:2020_Spin,Belczynski:2012_Mass-gap-lower}.

For all of the above reasons, it has now become commonplace to consider extended mass distributions. By far the most commonly considered case is the lognormal mass distribution, and constraints for this distribution were made by e.g.~\cite{Carr_2017,Bellomo_2018} (see \cite{Dolgov:1992pu} for the first related reference to this mass distribution in the PBH context). Broad mass distributions, such as a power law, or one with a spike at around one solar mass motivated by the QCD transition have also been considered~\cite{Carr:1975_Primordial,Byrnes:2018_QCD}, but in this paper we will focus on the more commonly studied case of a mass distribution generated by a single symmetric peak in the primordial power spectrum.

The lognormal mass distribution is frequently applied irrespective of its width, either in the form of priors allowing narrow widths (see e.g.~\cite{Hall:2020daa,Hutsi:2020sol,Franciolini:2021_Evidence}), or in the case of explicitly considering a very narrow case (see e.g.~\cite{Inomata:2020lmk}). However, it has been known for almost 25 years that for sufficiently narrow peaks in the power spectrum, the effects of critical collapse dominate. This creates a minimum width for the mass distribution, as discussed in~\cite{Gow:2020_power}; table~II in that work gives a minimum lognormal width of 0.37 based on a simple least squares fit to the numerical mass distribution calculated for a delta function peak in the power spectrum\footnote{A different value of 0.26 was stated in \cite{Carr_2017}, although this did not involve a full calculation of the mass distribution from a power spectrum peak, instead applying a method of moments approach to compare the lognormal with the critical collapse motivated shape in \cite{Yokoyama:1998xd}.}. Additionally, it is known that critical collapse causes the mass distribution shape to be significantly non-lognormal \cite{Niemeyer:1998_CC,Yokoyama:1998xd}. A large amount of work has been carried out on the mass distribution calculation since this deviation was first demonstrated, including the integration over all formation epochs mentioned but not pursued in \cite{Niemeyer:1998_CC} (\cite{Byrnes:2018_QCD}, see e.g.~\cite{Green:2001_Primordial,Green:2004_Peaks,Young:2020_Criterion,Suyama:2019_Novel,Germani:2020_Nonlinear,Young:2020_Peaks,Gow:2020_power} for further discussion of the calculation). This leads to two questions: does the conclusion of non-lognormality for narrow power spectrum peaks still hold and if so, is there a model for the PBH mass distribution that can accurately describe its behaviour for a broad range of power spectrum peak widths?

While the most rigorous choice is carrying out the full calculation of the mass distribution from the power spectrum, this can be computationally expensive, making it unsuitable for e.g.~Bayesian model selection calculations. Therefore, it is necessary to use models which allow an approximate capturing of the numerical mass distribution. Different constraints require the mass distribution to be narrow or broad, so it is essential to use a model that describes the numerical mass distribution for all these cases. In the following, we show that the lognormal assumption does indeed still break down for the narrowest widths, and propose some alternative models that can achieve a better fit over a large range of widths.

\section{The numerical mass distribution}
In order to test the validity of the lognormal mass distribution, we need a robust method of determining the PBH mass distribution corresponding to a particular peak in the primordial power spectrum. For this purpose, we use an accurate model for PBH formation described in Gow~\emph{et~al}~\cite{Gow:2020_power}. The procedure is to first relate the power spectrum peak to the PBH abundance $\Omega_{\PBH}(m)$, and then determine the mass distribution, given by
\begin{align}
\psi(m) &= \frac{1}{\Omega_{\PBH}}\ddv{\Omega_{\PBH}}{m}.
\end{align}
This is a probability distribution, and hence satisfies the condition $\int \diffd m\ \psi(m) = 1$, as will all the models we consider later.

The procedure to obtain the mass distribution is described in detail in section 2 of \cite{Gow:2020_power}. It incorporates the effects of critical collapse, and is robust to modelling choices at the 10\% level. In this paper, we will use the traditional peaks theory method with the modified Gaussian window function stated in eq.~(15) of \cite[arXiv version]{Gow:2020_power}. We additionally choose the same lognormal form for the primordial power spectrum peak,
\begin{align}
\mathcal{P}_\zeta = A\frac{1}{\sqrt{2\pi}\Delta}\exp\left(-\frac{\ln^2(k/k_p)}{2\Delta^2}\right), \label{eq:Ppeak}
\end{align}
which has a peak at $k_p$ and a width $\Delta$. The normalisation is chosen such that $\int \frac{\diffd k}{k}\ \mathcal{P}_\zeta(k) = A$, and means that this peak matches the case of a (Dirac) delta function $A\delta(\ln(k/k_p))$ in the limit $\Delta\to0$. The peak position is chosen such that the mass distribution peaks at $\sim35\ \Msun$. As noted in \cite{Gow:2020_power}, a broader power spectrum peak not only results in a broader mass distribution, but also a shift of the peak to lower masses. To ensure that the calculated mass distributions all peak at approximately the same mass, the position of the power spectrum peak is shifted accordingly. For the delta function case, $k_p=1.6\times10^6 \text{ Mpc}^{-1}$, corresponding to a horizon mass of $M_H = 7\ \Msun$. It should also be noted that in the LIGO mass range, there is an enhancement caused by the softening of the equation of state during the QCD phase transition \cite{Widerin:1998my,Jedamzik:1999am,Byrnes:2018_QCD,Carr:2019kxo}. We have neglected this effect so that the results regarding the optimal models are reliable at other mass scales. When considering a given mass range, any relevant thermal effects should be taken into account, which may mean that the models presented here need to be modified, but the relative ability of these models to fit the underlying distribution are not expected to change.

\begin{figure}[H]
\centering
\includegraphics[width=0.8\textwidth]{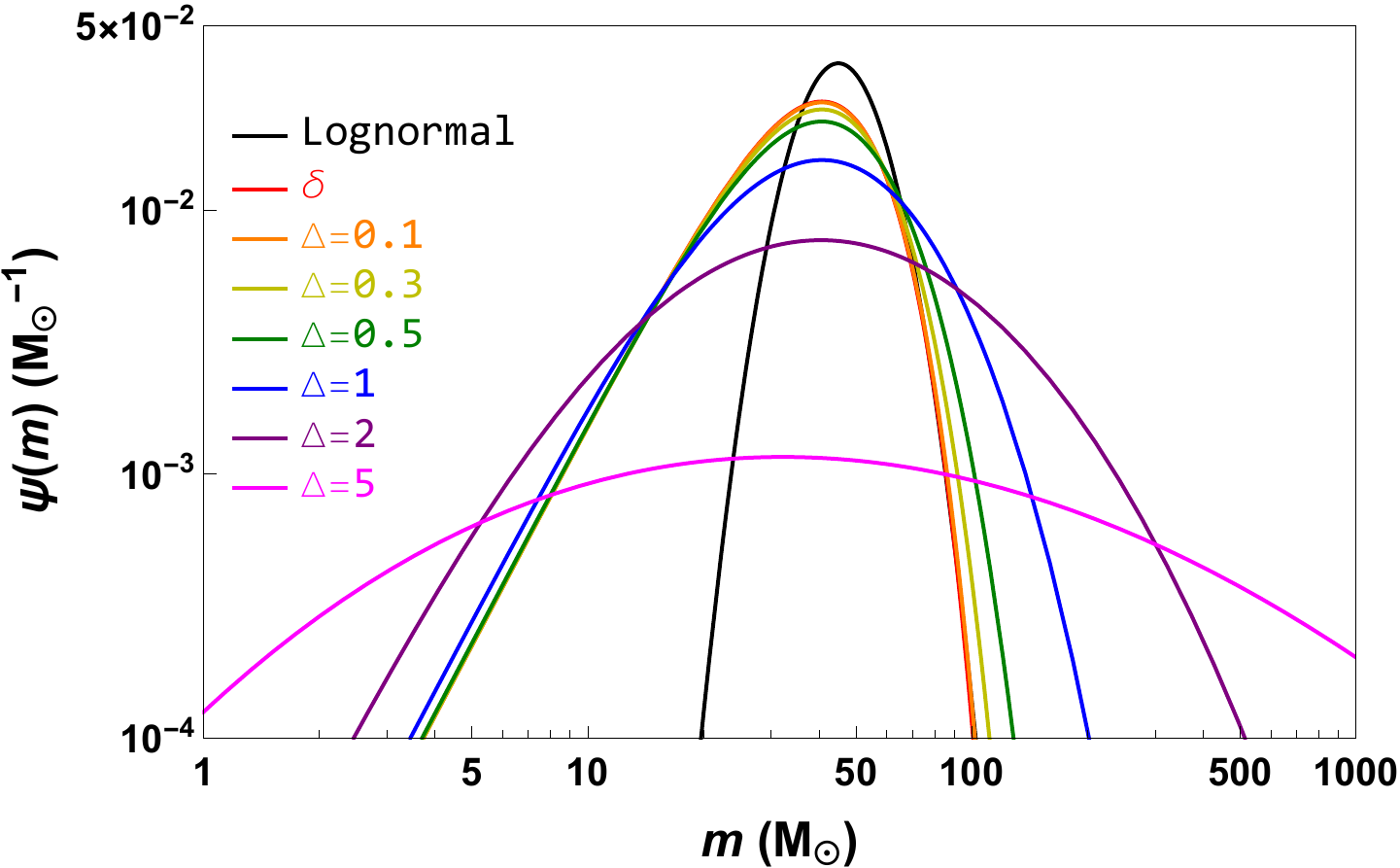}
\caption{The numerical mass distribution calculated for a range of power spectrum peak widths. The peak positions are chosen for each width such that the resulting distribution peaks at $\sim35\ \Msun$. The black curve shows a representative lognormal with a peak mass and width chosen such that its high-mass tail coincides with that of the narrowest numerical mass distributions.}
\label{fig:psiPlot_num}
\end{figure}

We can see from fig.~\ref{fig:psiPlot_num} that, taking into account the changes to the mass distribution calculation over the last 25 years, the non-lognormality seen in \cite{Niemeyer:1998_CC,Yokoyama:1998xd} is still valid, with the narrowest peaks showing significant deviation from the symmetric shape expected for a lognormal mass distribution.

\subsection{Narrow peaks and critical-collapse domination}
\begin{figure}[H]
\centering
\includegraphics[width=0.49\textwidth]{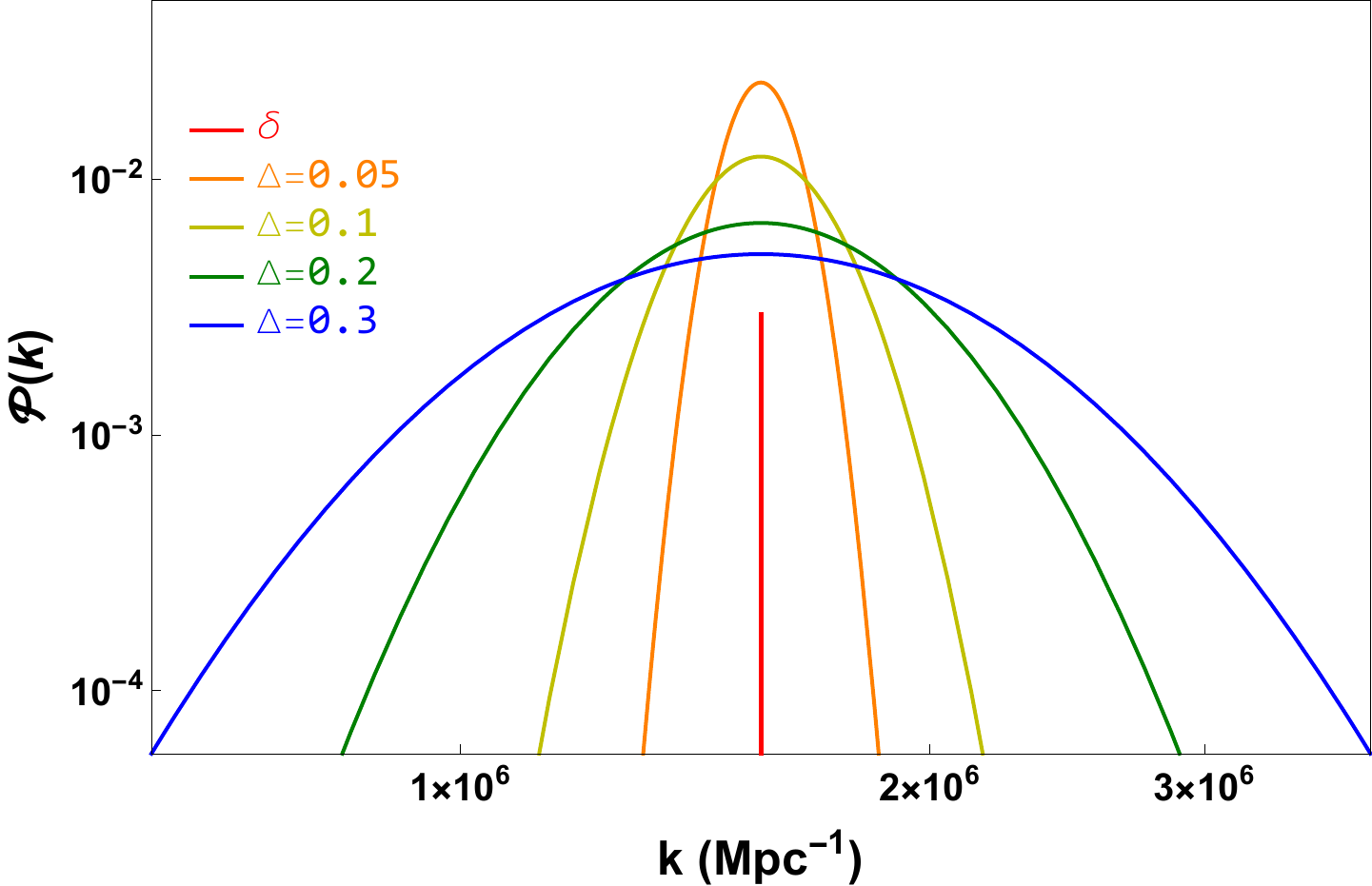}\hspace{0.01\textwidth}\includegraphics[width=0.49\textwidth]{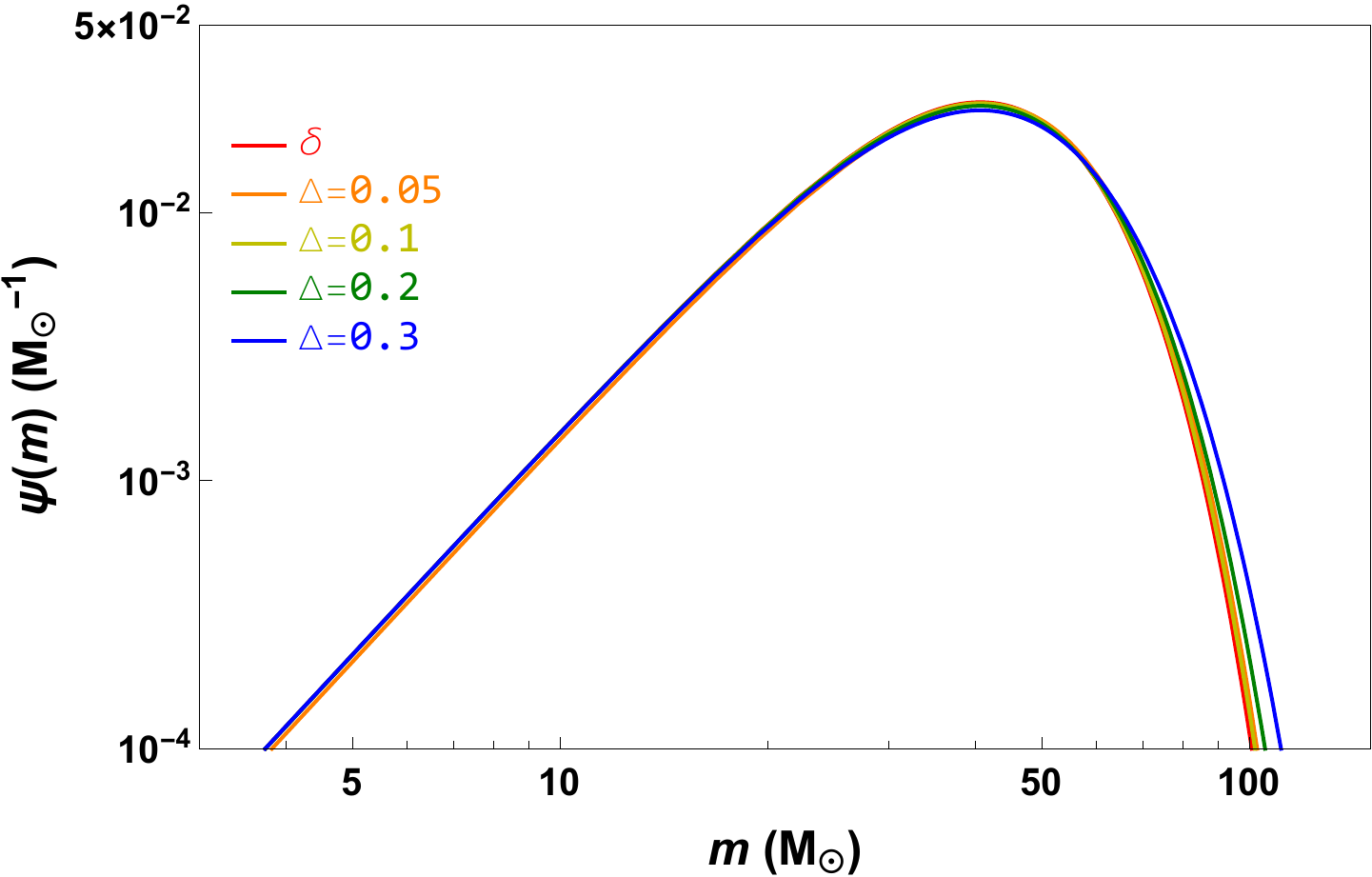}
\caption{Narrow lognormal peaks in the power spectrum (left) and the PBH mass distributions they generate (right). The power spectrum peaks have amplitudes chosen to produce the same $f_\PBH$, and their locations coincide to highlight the variation in their widths. The mass distributions are constructed to peak at $\sim35\Msun$.}
\label{fig:psiPlot_num-narrow}
\end{figure}

Figure~\ref{fig:psiPlot_num} shows that a significant asymmetry arises from the effects of critical collapse for narrow peaks in the primordial power spectrum. Additionally, we can see that the three narrowest mass distributions look virtually identical, indicating that below a certain width, the shape of the mass distribution is dominated by critical collapse and is insensitive to the details of the power spectrum peak. We can examine this in more detail by considering a larger number of narrow cases for the lognormal. This is done in fig.~\ref{fig:psiPlot_num-narrow}, where we show the power spectra on the left and the corresponding mass distributions on the right.

It is immediately obvious that despite a significant change in the power spectrum peak, the resulting mass distribution remains essentially unchanged, confirming that the shape is being set by critical collapse rather than the power spectrum peak. This implies two interesting conclusions. First, since the mass distribution is dominated by critical collapse effects, we expect the shape to remain identical even for significantly different power spectrum peaks, such as a power-law or a Gaussian, provided that they are sufficiently narrow. Secondly, and following from this first conclusion, it may be impossible to completely describe a power spectrum peak that generates primordial black holes, even if the abundance and mass distribution are measured with infinite precision. Instead, the mass distribution shape will only be able to confirm the physics of critical collapse. Nonetheless, it is important to have a model that can describe this critical collapse shape as well as the broader cases where the peak shape does become important, such that a general fit can be made that can discriminate between these different cases.

\section{Modelling the PBH mass distribution}
\subsection{Fitting procedure}
To find the best fitting mass distribution model, we use a $\chi^2$ statistic,
\begin{align}
\chi^2 &= \frac{1}{\psi_\text{peak}^2}\sum_i [\psi_\text{num}(m_i)-\psi_\text{model}(m_i)]^2 w_i, \label{eq:chiSq}
\end{align}
with the weightings given by
\begin{align}
w_i &= \left(\frac{\psi_\text{num}(m_i)}{\psi_\text{peak}}\right)^2.
\end{align}
We make this choice such that the peak receives more weight than the tails, since the majority of observational techniques are most sensitive to the masses around the peak. The squared power is motivated by considering a fit to the LIGO merger rate data, where the merger rate is (roughly) proportional to $\psi^2$. The overall normalisation by $\psi_\text{peak}^2$ is similar to fitting $\psi/\psi_\text{peak}$, and ensures that the $\chi^2$ values can be compared not only between models, but also for the same model with different widths.

The data are drawn from the numerical mass distribution, and consist of 100 values spaced equally in log mass. We choose a log-spacing because the constraints on the mass distribution stretch over many orders of magnitude, and the low-mass tail must be fitted with comparable weight to the high-mass tail in order to ensure that relevant constraints are not missed. The lower mass tail is especially relevant for e.g.~microlensing constraints if we want a peak in the LIGO range, or evaporation constraints for a peak in the asteroid mass band. The limits are set arbitrarily to encompass the top four orders of magnitude of the distribution. The weighting applied to the $\chi^2$ statistic should mean that any part of the mass distribution outside of these limits will contribute negligibly to the best fit.

\subsection{Models}
In this section, we present various parametrisations considered for the PBH mass distribution.
\subsubsection{Lognormal}
The de-facto standard mass distribution considered for PBHs generated from a reasonably narrow, smooth, symmetric peak in the power spectrum is the lognormal, given by
\begin{align}
\psi_\text{L}(m) &= \frac{1}{\sqrt{2\pi}\sigma m}\exp\left(-\frac{\ln^2(m/m_c)}{2\sigma^2}\right),
\end{align}
where $m_c$ is the mean of $m\psi(m)$ and $\sigma$ is the width. There are a number of alternative distributions to the lognormal that may fit the numerical mass distribution better over the whole range of power spectrum peak widths. The ones chosen for testing in this work are described in the following sections.

\subsubsection{Gaussian}
This is simply a standard Gaussian distribution, given by
\begin{align}
\psi_\text{G}(m) &= \frac{1}{\sqrt{2\pi}\sigma}\exp\left(-\frac{\left(m-m_c\right)^2}{2\sigma^2}\right),
\end{align}
with $m_c$ the mean and $\sigma$ the width. It should be noted that this distribution allows for negative masses, which are clearly unphysical. However, if the fit is good, the fraction of negative masses should be negligible.

\subsubsection{Skew-normal}
The skew-normal is a modification to the Gaussian distribution which introduces skewness by multiplying the Gaussian PDF with a Gaussian CDF modified with a parameter $\alpha$. The definition is
\begin{align}
\psi_\text{SN}(m) &= \frac{1}{\sqrt{2\pi}\sigma}\exp\left(-\frac{(m-m_c)^2}{2\sigma^2}\right)\left[1 + \text{erf}\left(\alpha\frac{m - m_c}{\sqrt{2}\sigma}\right)\right].
\end{align}
As for the Gaussian, this distribution can produce negative masses, although the fraction is expected to be small for a good fit.

\subsubsection{Skew-lognormal}
The skew-lognormal is virtually identical to the skew-normal, but with the mass terms switched for log-mass terms, and an additional factor of $1/m$ to preserve the normalisation over mass, i.e.,
\begin{align}
\psi_\text{SL}(m) &= \frac{1}{\sqrt{2\pi}\sigma m}\exp\left(-\frac{\ln^2(m/m_c)}{2\sigma^2}\right)\left[1 + \text{erf}\left(\alpha\frac{\ln(m/m_c)}{\sqrt{2}\sigma}\right)\right].
\end{align}
It can be seen that, excluding the last bracket, this is simply the lognormal mass distribution, hence the name skew-lognormal. Since this is defined in log-space, it is superior to the skew-normal in that it avoids producing negative masses.

\subsubsection{Critical collapse models}
Motivated by the mass distribution dominated by critical collapse effects calculated in \cite{Niemeyer:1998_CC,Yokoyama:1998xd}, and later models based upon this form \cite{Carr_2017,Vaskonen:2021_NANOGrav}, we introduce a critical collapse model class, given in general by
\begin{align}
\psi_\text{CC}(m) &= \frac{\beta}{m_f} \left[\Gamma\left(\frac{\alpha+1}{\beta}\right)\right]^{-1}\left(\frac{m}{m_f}\right)^\alpha \exp\left[-\left(\frac{m}{m_f}\right)^\beta\right],
\end{align}
where the PBH mass is given by the critical collapse equation
\begin{align}
m &= KM_H(\delta-\delta_c)^\gamma,
\end{align}
where $M_H$ is the horizon mass at formation, $K$ is a dimensionless constant, $\gamma\simeq0.36$ is a universal scaling exponent which is independent of the initial shape of the density fluctuations and $\delta_c$ is the minimum overdensity required for PBH formation \cite{Musco:2009_CC}. In this class, we consider three models, defined as follows:
\begin{itemize}
\item CC1: $\alpha=\beta=\gamma^{-1}$, $\gamma = 0.36$

This is the most simple model, motivated entirely by the critical collapse calculations. It is identical to the form stated in~\cite{Yokoyama:1998xd}, and has been numerically checked for small $\delta-\delta_c$ in~\cite{Musco:2009_CC}. It has just one parameter, to fit the location of the distribution.
\item CC2: $\alpha=\beta=\gamma^{-1}$, $\gamma$ variable

The shape of this model is identical to the above case, in that both tails are described by $\gamma$. However, in this case, we allow $\gamma$ to float to find the best fit. This is motivated by the demonstration in \cite{Escriva:2020_Simulation} that the value $\gamma=0.36$ does not hold for larger values of $\delta-\delta_c$, and a modification to the critical collapse parameters $K$ and $\gamma$ could yield a better fit across the whole range. This model has two parameters, to fit the location and shape of the distribution.
\item CC3: $\alpha$, $\beta$ variable

This is a generalisation of the critical collapse model, disconnecting the behaviour of the two tails. It has three parameters, to fit the location and the shape of each tail.
\end{itemize}

\subsubsection{Location parameter}
The location parameters stated in the parametrisations above can be extremely sensitive to the width of the mass distribution, causing problems in the fitting procedure. To overcome this, we reparametrise most of the models in terms of their peak mass $m_p$, which we have held approximately fixed for all of the numerical mass distributions. The transformations between the peak mass and the location parameters defined above are given in table~\ref{tab:location_parameters}.

For the skew-normal and skew-lognormal there is no analytical form for the peak mass. There is an approximate transformation derived from numerical fits \cite{Azzalini:2013_Skew-normal}, but this does not hold for the broadest cases. Therefore, for the skew-normal and skew-lognormal we retain the location parameters $m_c$ and $\ln(m_c)$ defined above. For these two models these location parameters are stable enough to avoid numerical errors during fitting.

\begin{table}[H]
\centering
\caption{Transformation of location parameter to peak mass $m_p$ for each model.}
\label{tab:location_parameters}
\begin{tabular}{c|c}
\textbf{Model} & \textbf{Transformation} \\ \hline
Lognormal & $m_c = m_pe^{\sigma^2}$ \\
Gaussian & $m_c = m_p$ \\
Skew-normal & N/A \\
Skew-lognormal & N/A \\
CC1\&2 & $m_f=m_p$ \\
CC3 & $m_f = m_p\left(\frac{\beta}{\alpha}\right)^{1/\beta}$
\end{tabular}
\end{table}

\subsection{Fit results}
\begin{figure}[H]
\centering
\includegraphics[width=0.5\textwidth]{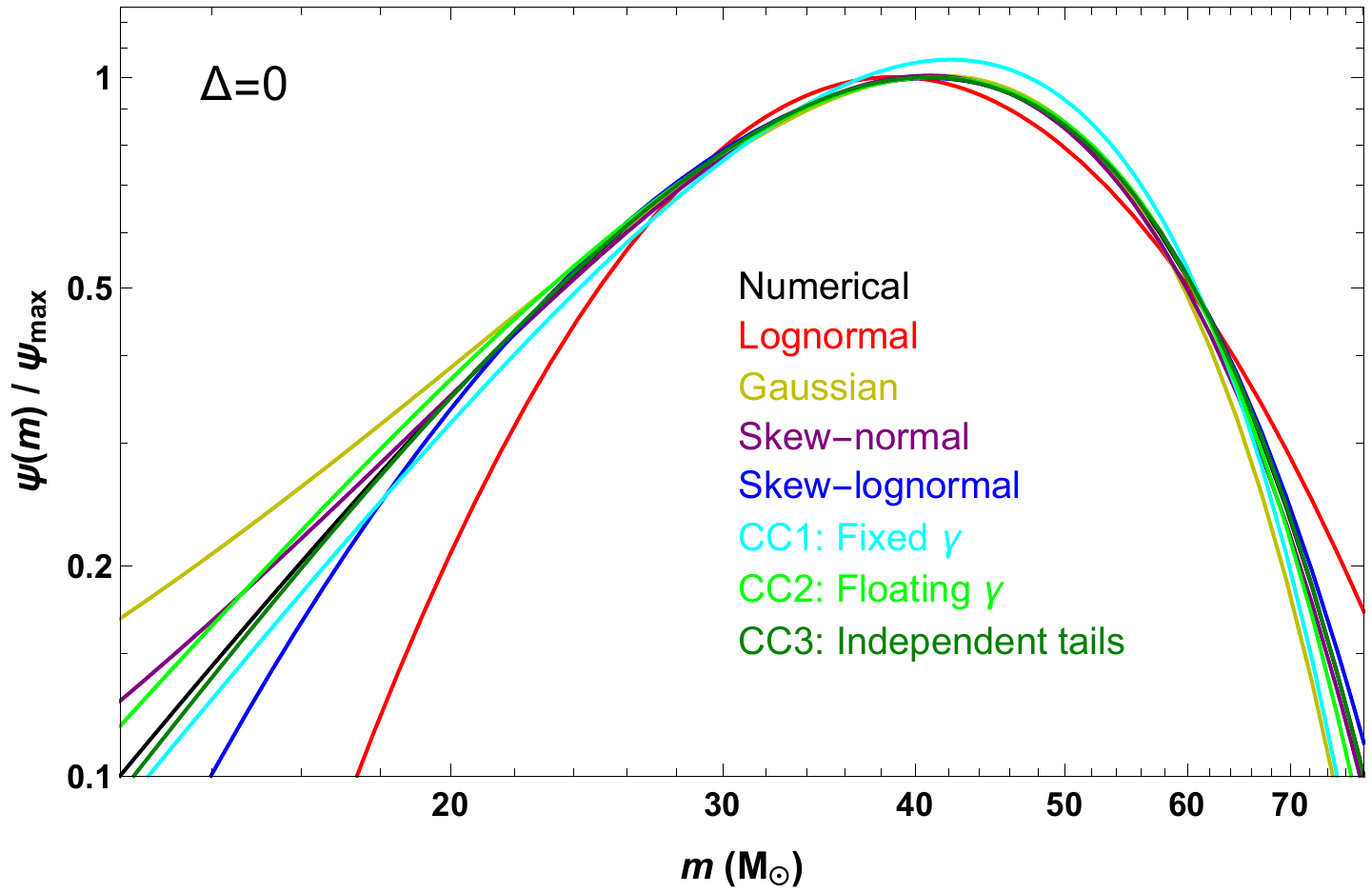}\includegraphics[width=0.5\textwidth]{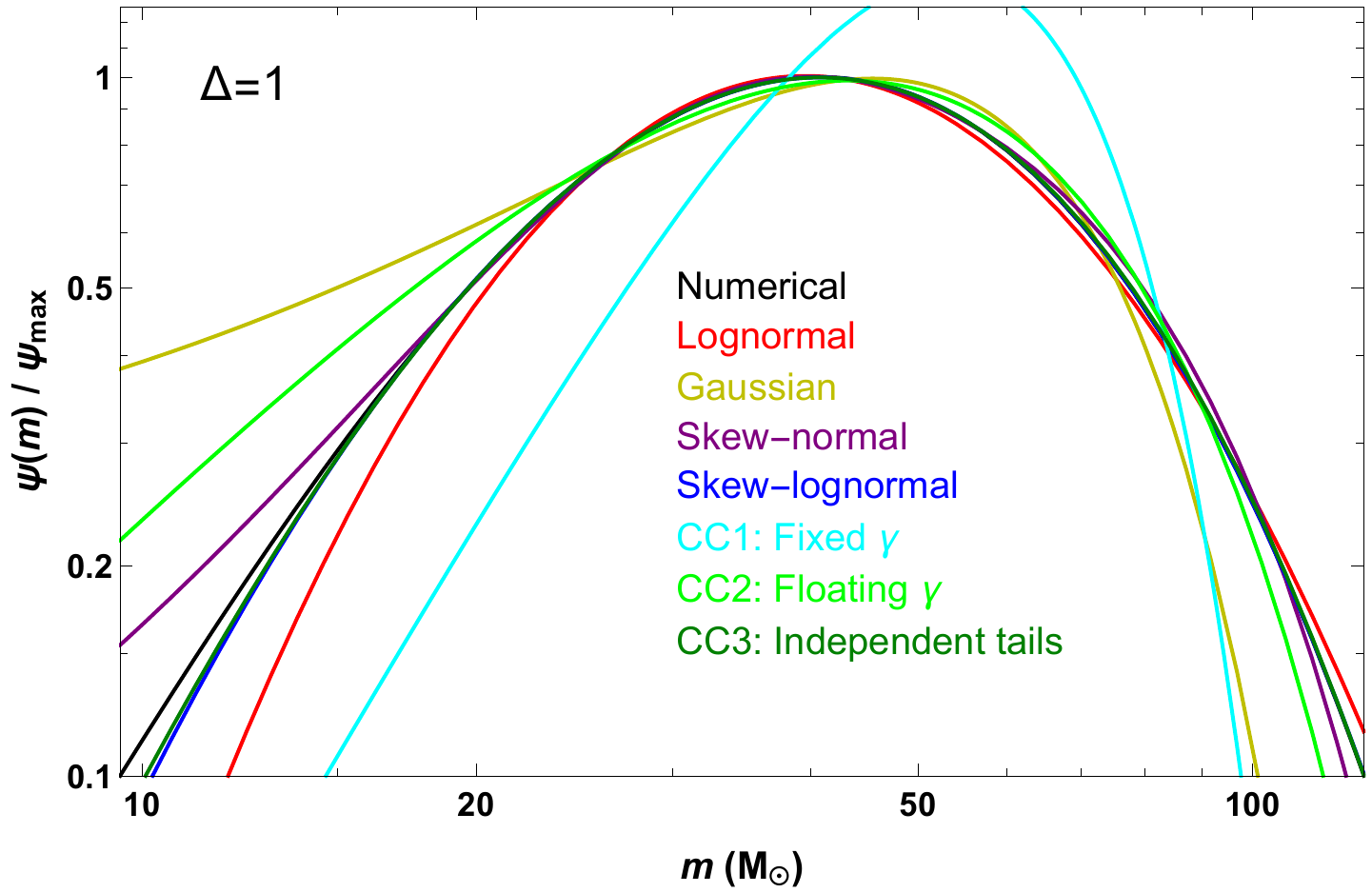}
\includegraphics[width=0.5\textwidth]{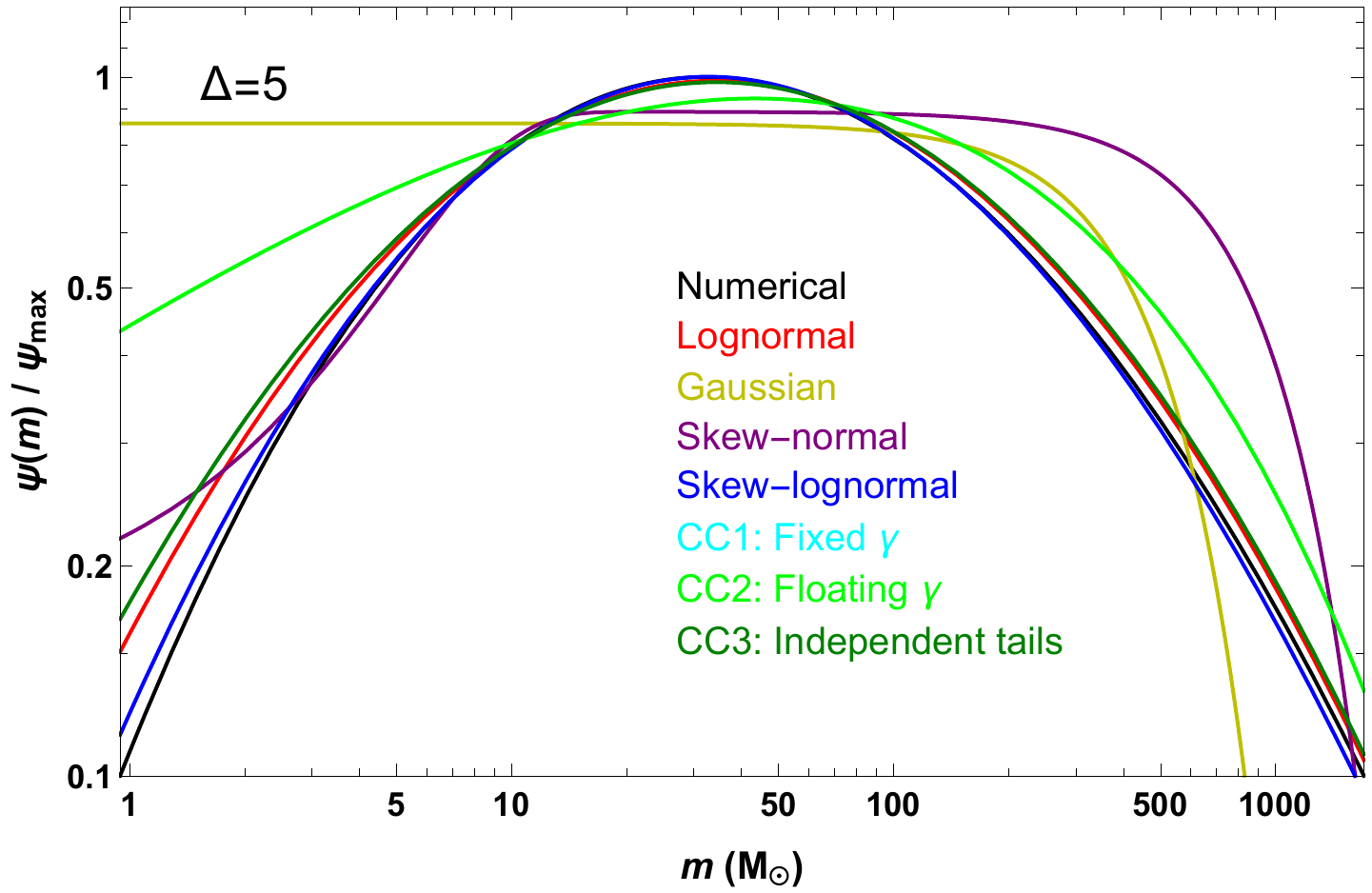}
\caption{Optimal model fits to the numerical mass distribution for three representative power spectrum widths $\Delta=0$ (delta function), $\Delta=1$, and $\Delta=5$. The mass limits are chosen to contain the top 10\% of the numerical mass distribution, to highlight the deviation of the models near the peak.}
\label{fig:psiPlot_fit-all}
\end{figure}

We obtain fits to the numerical mass distribution calculated from a peak in the power spectrum by minimisation of eq.~\eqref{eq:chiSq}. We consider a large range of power spectrum peak widths, from the limiting case of a delta function up to a very broad case of $\Delta=5$. The optimised model fits are shown in fig.~\ref{fig:psiPlot_fit-all} for three representative cases: a delta function, $\Delta=1$, and $\Delta=5$. It is immediately apparent that the lognormal is outperformed by the vast majority of the models for the narrowest case. However, it can be seen that many of these models begin to fail as the width increases, and are completely wrong for the broadest case.

\begin{figure}[H]
\centering
\includegraphics[width=0.76\textwidth]{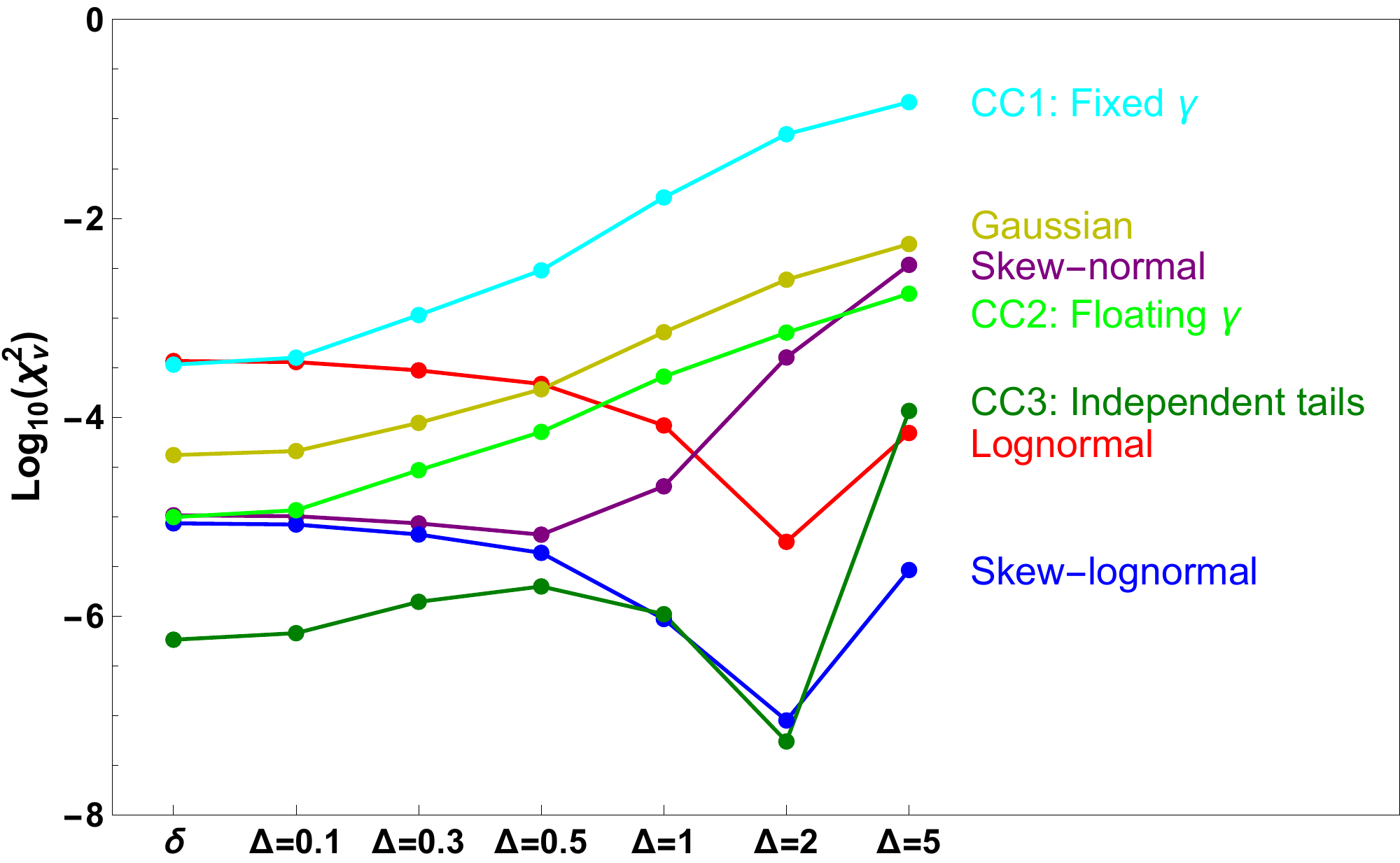}
\caption{Reduced $\chi^2$ values for the models and widths considered. Lower (more negative) values indicate a better fit.}
\label{fig:chiSqPlot}
\end{figure}

We can compare the models more carefully by examining their reduced $\chi^2$ values. Figure~\ref{fig:chiSqPlot} shows the $\chi^2_\nu$ values for all the models and widths considered. Here we can see again that, while there are many models that outperform the lognormal for the narrowest cases, a large number of them fail as the width increases, where they cannot generate the appropriate skewness. However, it can be seen that two models, the skew-lognormal and the generalised critical collapse model, consistently provide a more accurate fit than the lognormal. These models also have the benefit of not producing negative masses, although the models which do allow this are deemed irrelevant by their failure to fit the broadest cases anyway. The reduced $\chi^2$ values are provided in table~\ref{tab:stat_values}.

\begin{figure}[H]
\centering
\includegraphics[width=0.47\textwidth]{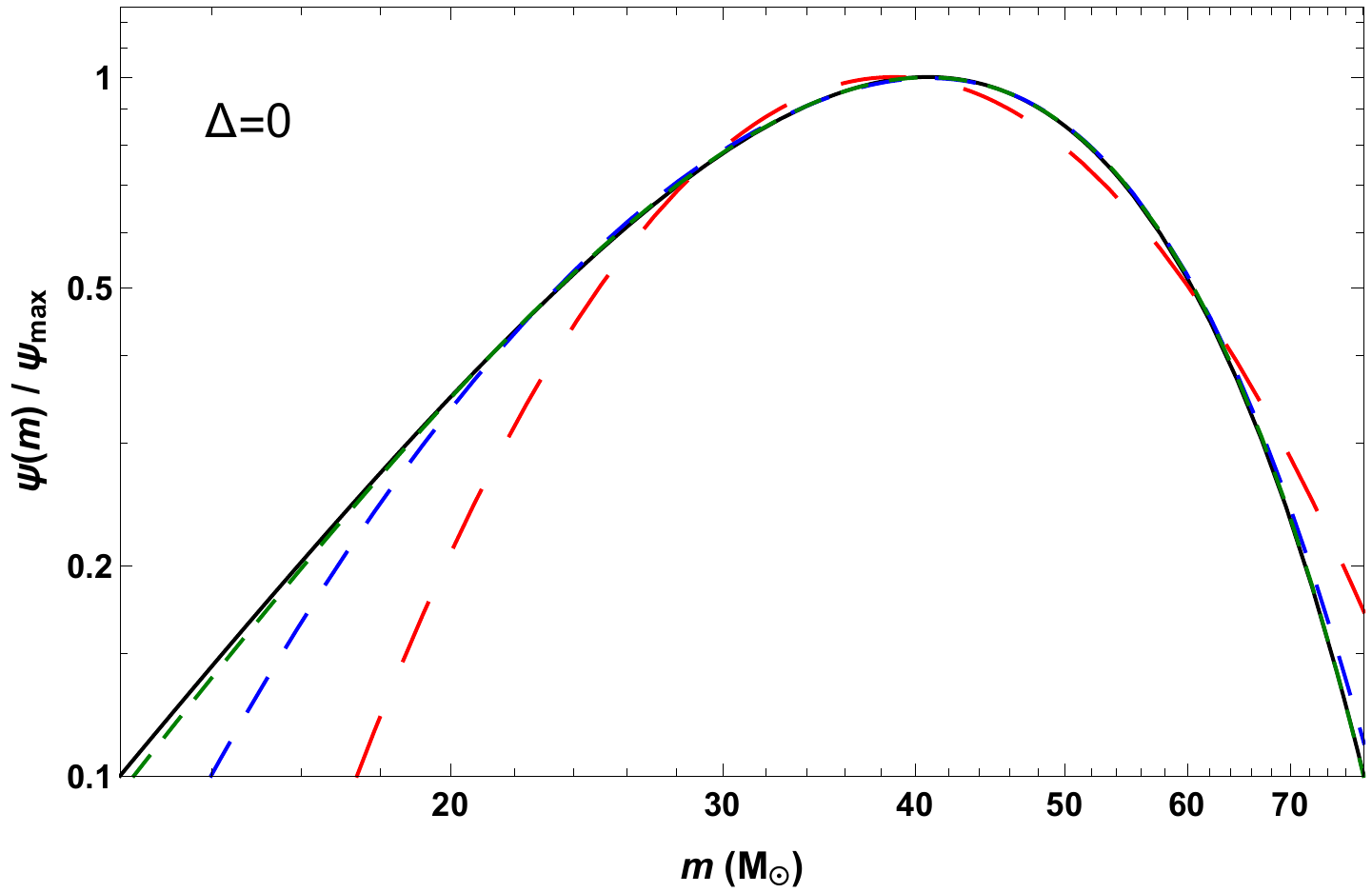}
\includegraphics[width=0.47\textwidth]{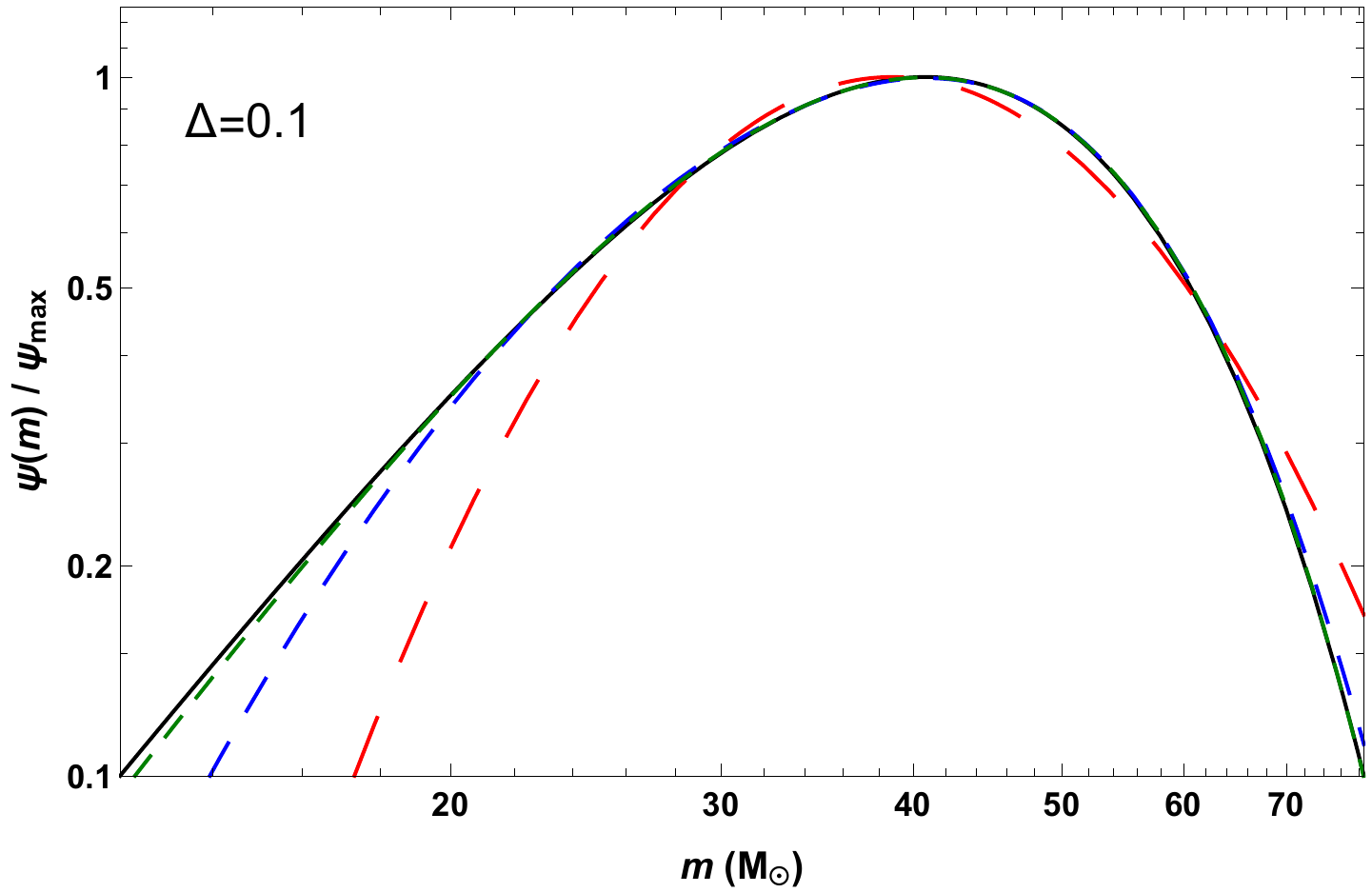}\includegraphics[width=0.47\textwidth]{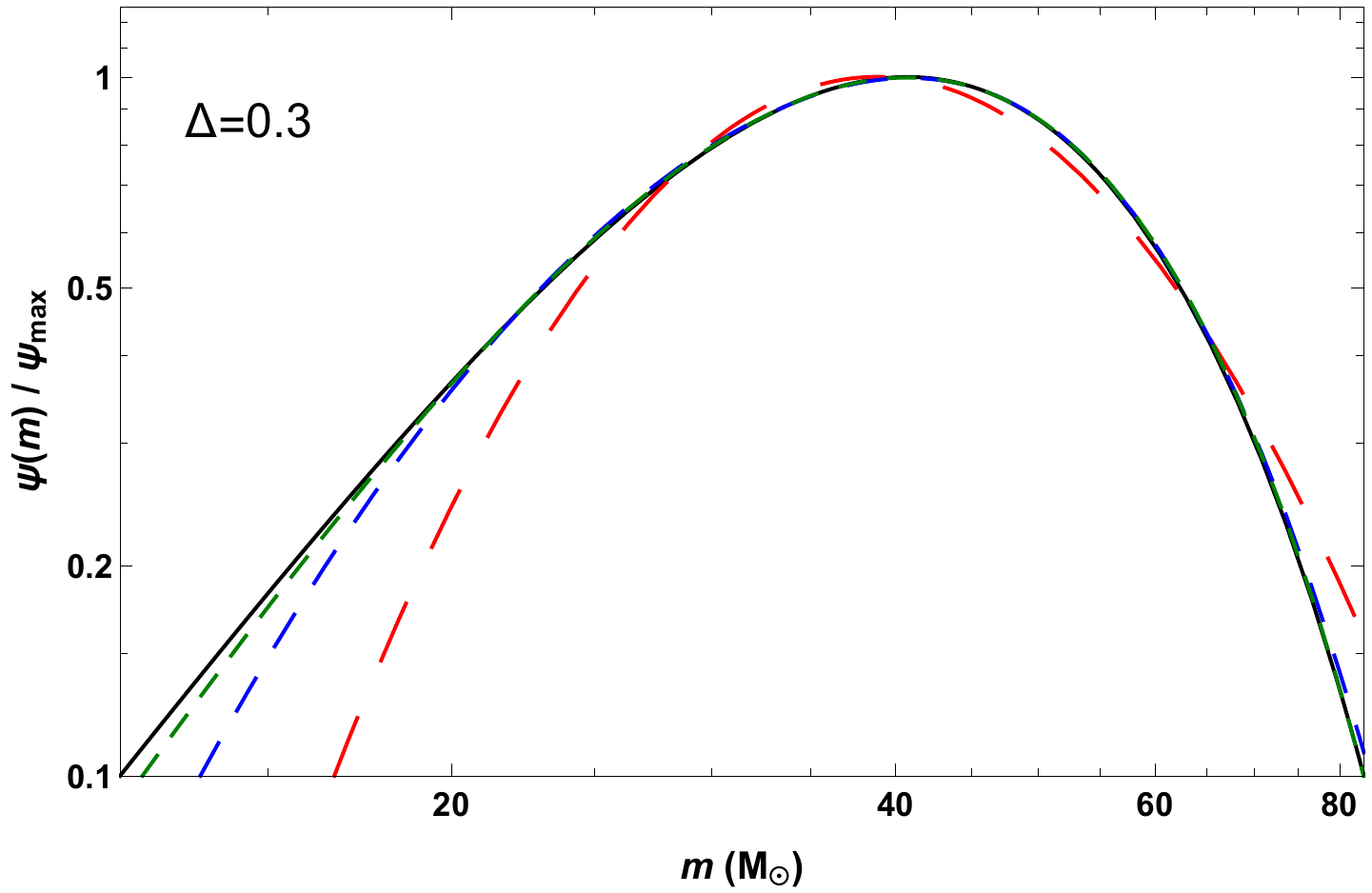}
\includegraphics[width=0.47\textwidth]{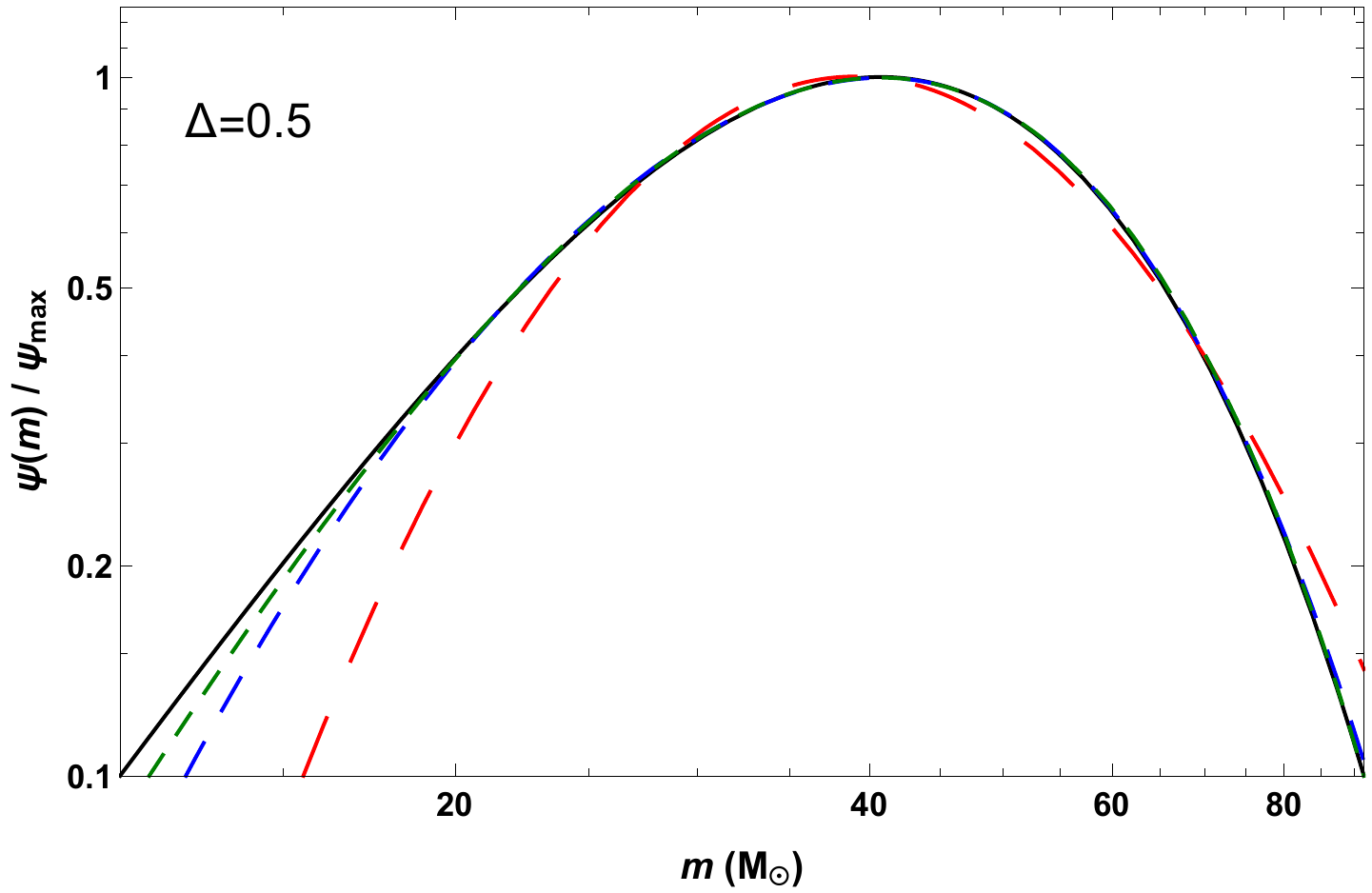}\includegraphics[width=0.47\textwidth]{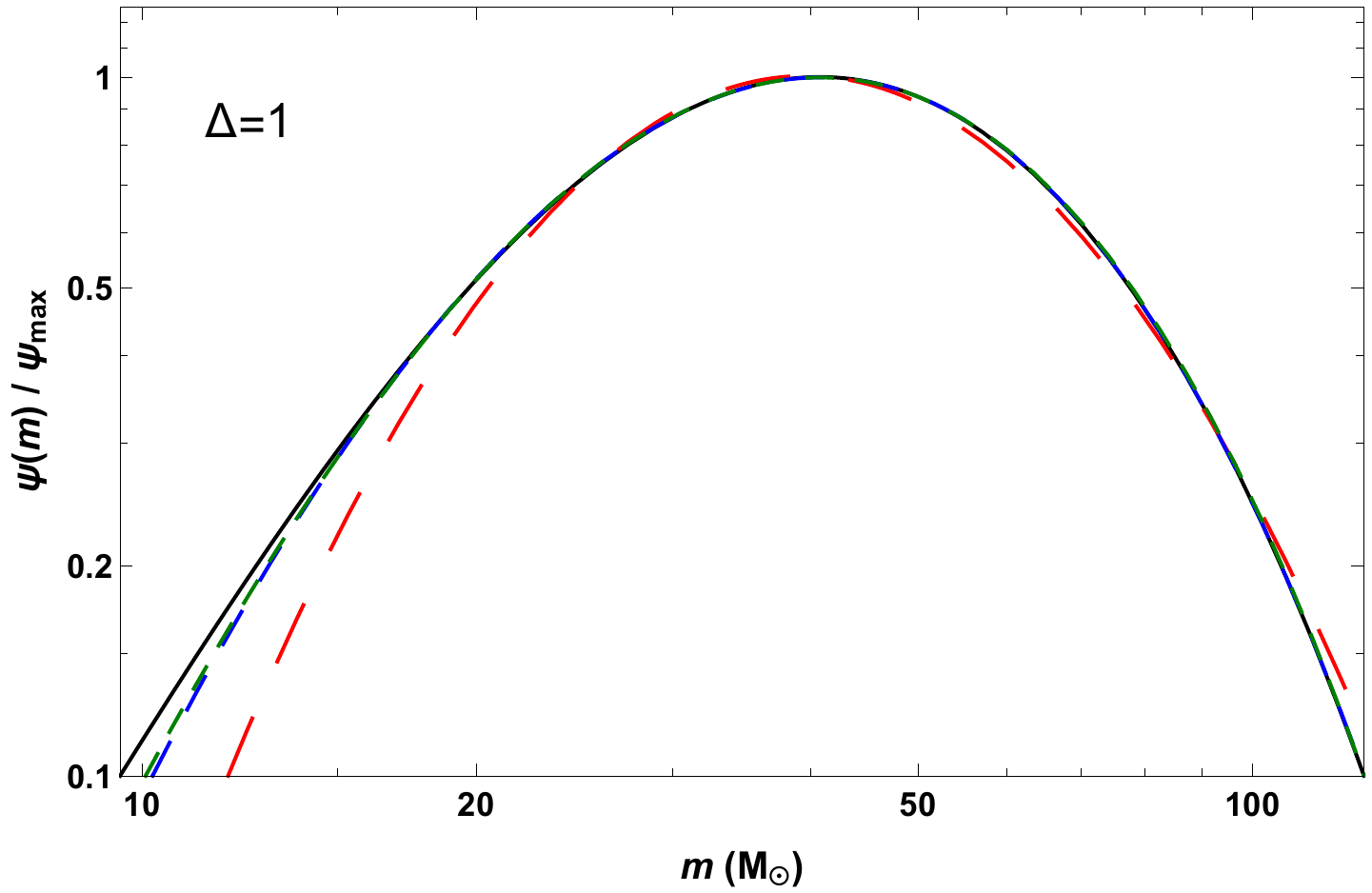}
\includegraphics[width=0.47\textwidth]{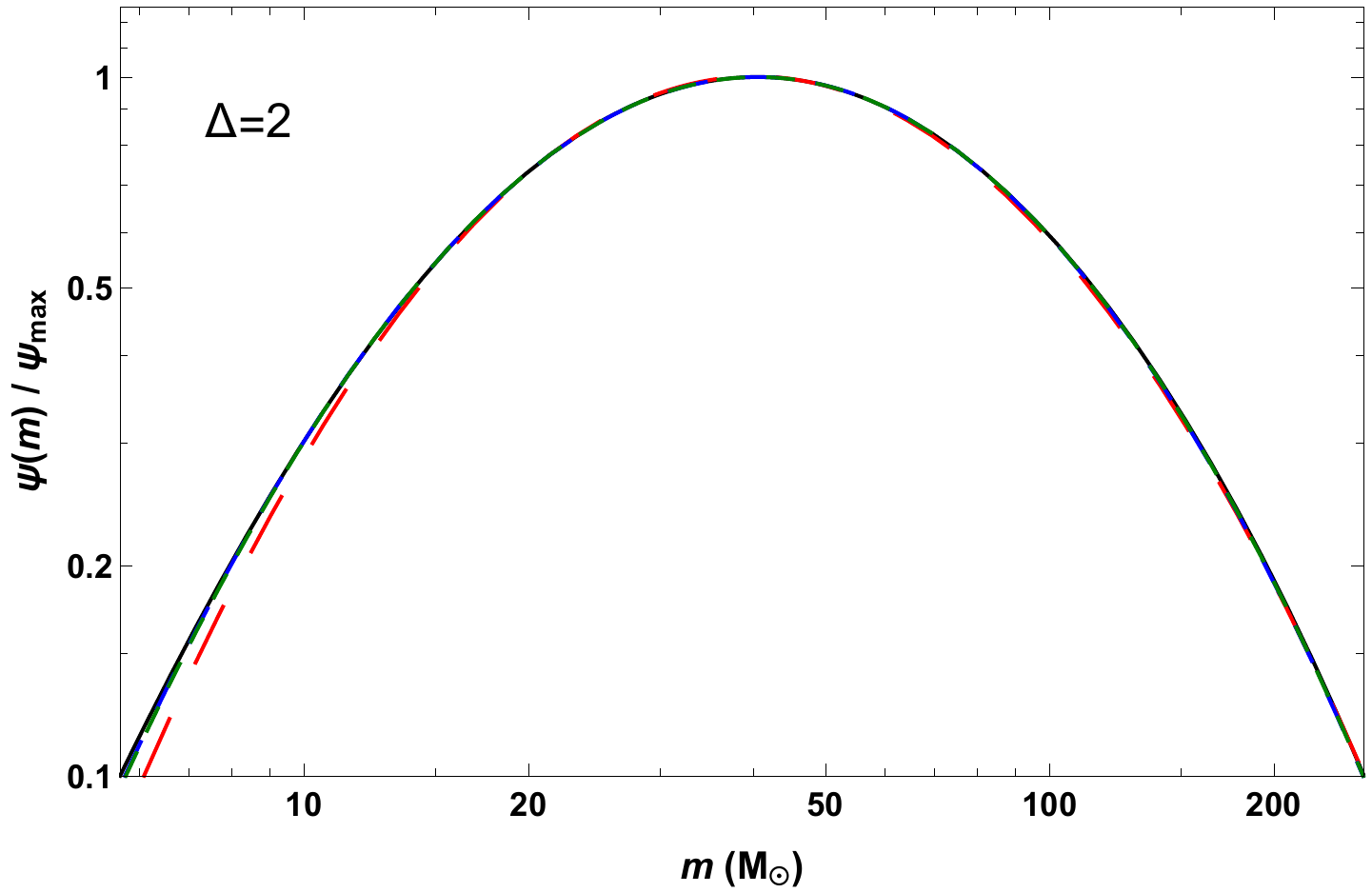}\includegraphics[width=0.47\textwidth]{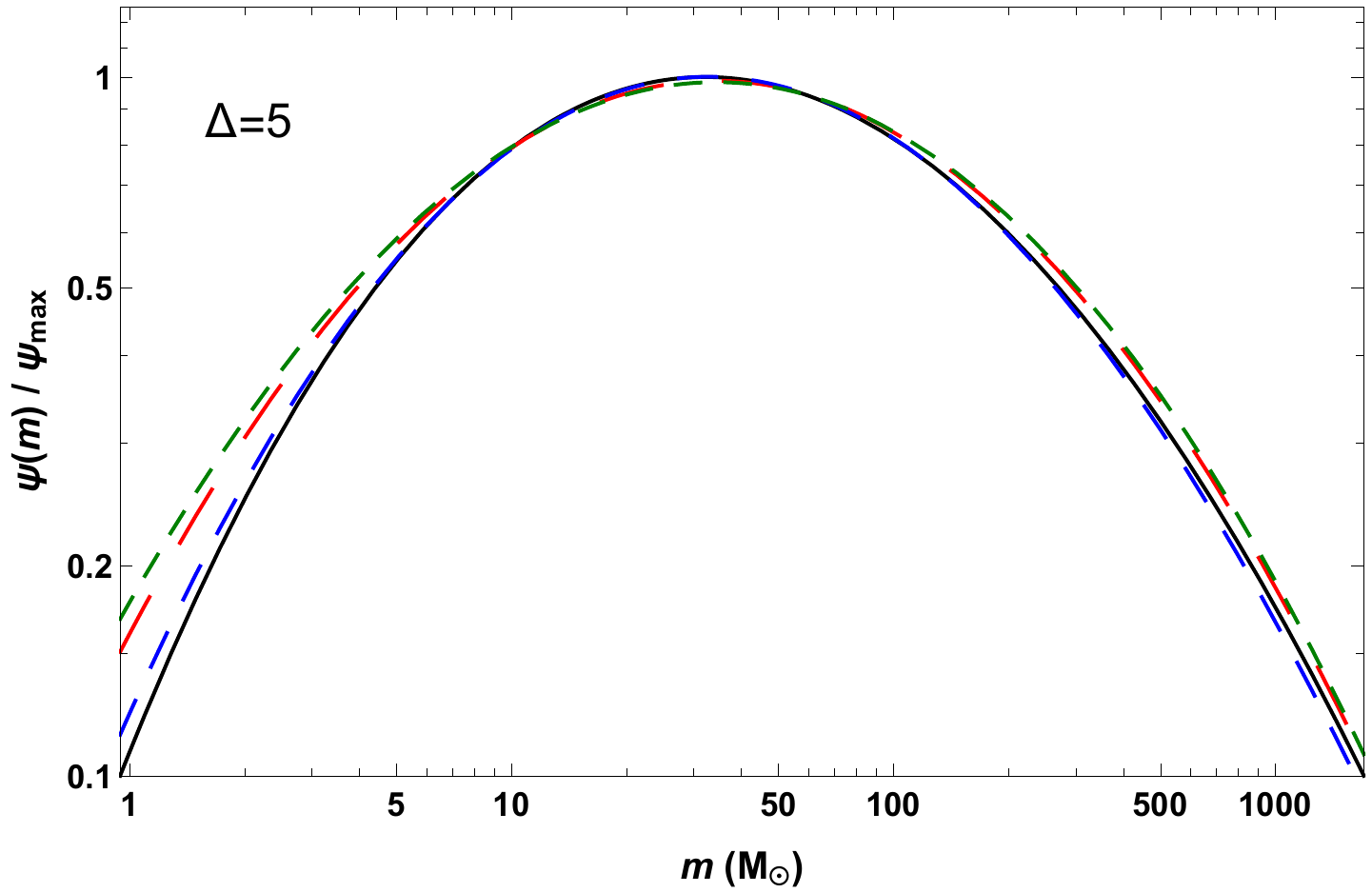}
\caption{Plots of the lognormal (red, long-dashed), skew-lognormal (blue, mid-dashed) and generalised critical collapse (green, short-dashed) fits to the numerical mass distribution generated by a lognormal peak in the power spectrum (black, solid). The mass limits are chosen to contain the top 10\% of the numerical mass distribution, to highlight the deviation of the models near the peak.}
\label{fig:psiPlot_LN-best}
\end{figure}

The comparison between the lognormal and the two models that consistently outperform it can be seen graphically in fig.~\ref{fig:psiPlot_LN-best}, where the best fit lognormal is shown with a long-dashed red line, the skew-lognormal with a mid-dashed blue line, and the generalised critical collapse model with a short-dashed green line. The numerical distribution calculated from the power spectrum is shown with a solid black line.

It is evident from these plots that modelling the PBH mass distribution across a broad range of widths is a challenging task, as it requires negative skewness in log-space for the narrowest cases, before a change to symmetrical and then positively skewed distributions. The model best suited for the job is the generalised critical collapse model, which can produce the negative skewness exceptionally well, but begins to fail when positive skewness is required. The skew-lognormal acts oppositely, with a failure to produce enough negative skewness, but an improvement for the positive skewness regime.

\begin{table}[H]
\centering
\caption{Fitted parameter values for the skew-lognormal and generalised critical collapse distributions with different power spectrum widths. For the skew-lognormal model, we also provide the peak mass $m_p$ determined by numerical maximisation. It should be noted that $\ln(m_c)$ and $m_p$ are not independent and only $\ln(m_c)$ is determined by the fit for the skew-lognormal model. The peak mass is included only for comparison to the fitted parameter in the critical collapse model.}
\label{tab:Best_model_fit_params}
\begin{tabular}{c|cccc|ccc}
& \multicolumn{7}{c}{\textbf{Parameters}} \\
& \multicolumn{4}{c}{\textbf{SL}} & \multicolumn{3}{|c}{\textbf{CC3}} \\
\textbf{Width} $\boldsymbol{\Delta}$ & $\boldsymbol{\ln(m_c)}$ & $\boldsymbol{\sigma}$ & $\boldsymbol{\alpha}$ & $\boldsymbol{m_p}$ & $\boldsymbol{m_p}$ & $\boldsymbol{\alpha}$ & $\boldsymbol{\beta}$ \\ \hline
$\delta$ & 4.13 & 0.55 & $-2.27$ & 40.9 & 40.8 & 3.06 & 2.12 \\
$0.1$ & 4.13 & 0.55 & $-2.24$ & 40.9 & 40.8 & 3.09 & 2.08 \\
$0.3$ & 4.15 & 0.57 & $-2.07$ & 40.9 & 40.7 & 3.34 & 1.72 \\
$0.5$ & 4.21 & 0.60 & $-1.82$ & 40.8 & 40.7 & 3.82 & 1.27 \\
$1.0$ & 4.40 & 0.71 & $-1.31$ & 40.8 & 40.8 & 5.76 & 0.51 \\
$2.0$ & 4.88 & 0.97 & $-0.66$ & 40.6 & 40.6 & 18.9 & 0.0669 \\
$5.0$ & 5.41 & 2.77 & $\hphantom{-}1.39$ & 32.9 & 35.1 & 13.9 & 0.0206 \\ \hline
\end{tabular}
\end{table}

Of course, there is a price to pay for achieving this matching, and that is the introduction of an additional parameter. Both of the best-fitting models have three parameters, as opposed to the two required for the lognormal. However, the importance of accurately describing the shape of the tails of the distribution cannot be overstated. Failure to capture this shape can result in incorrect conclusions about the acceptability of a particular model. For example, if we are looking for PBHs in the LIGO mass range, we must ensure that the tails of the distribution do not conflict with the microlensing constraints on the low-mass side and the CMB anisotropy constraints on the high-mass end. Similarly, for PBHs in the recently reopened asteroid mass window, a significant underfitting of the low-mass tail, such as that displayed by the lognormal model for the narrower widths, could suggest that PBHs can evade all the constraints, whereas a more accurate model would show that they are in tension with the evaporation limits. In table~\ref{tab:Best_model_fit_params}, we provide the optimal model parameters for the best two models, the skew-lognormal and the generalised critical collapse model, for the widths considered. This allows fits of these analytical approximations to be crudely compared to the power spectrum details without the necessity of recalculating the full mass distribution. However, it should be noted that although these more accurate models provide a significant improvement over the lognormal, even they fail to capture the detailed shape deep into the tails, and the only truly rigorous way to determine whether PBHs are not excluded in a given range is to calculate the mass distribution from the power spectrum peak.

\section{Conclusions}
We have carried out a thorough examination of the PBH mass distribution arising from a peak in the primordial power spectrum, re-evaluating the validity of the lognormal approximation to the mass distribution. We confirm that the modifications to the PBH mass distribution calculation over the last 25 years do not change the conclusion that the lognormal model is still unable to accurately capture the shape of the distribution generated from sufficiently narrow peaks, with $\Delta<1$. We compare a set of alternative models using a weighted $\chi^2$ statistic, and show that over a large range of peak widths, the lognormal is outperformed by the skew-lognormal and a generalised form motivated by the effects of critical collapse.

This deviation between the lognormal assumption and the PBH mass distribution calculated for a specific power spectrum peak will have important consequences for physical inferences made from accurate data, such as the LIGO--Virgo observations. In a related previous paper \cite{Hall:2020daa}, we considered the skew-lognormal as part of a detailed Bayesian analysis of the LIGO--Virgo O1O2 dataset. The limited sample size means that the difference in the mass distribution does not significantly affect the results, but the difference will become increasingly important with the accurate data in the O3 run and future runs.

An accurate model of the PBH mass distribution will also be relevant in other areas, such as making accurate constraints on the PBH abundance. These constraints are typically presented for a monochromatic mass distribution, but extended mass distributions have been considered, see e.g.~\cite{Carr_2017,Bellomo_2018}. The constraints for extended mass distributions are typically similar to the monochromatic case, but the differences become important when determining the validity of specific extended mass distributions, particularly in the case of $f_\PBH\sim1$, where the tails of the distributions may be in tension with constraints. In these cases, an accurate model of the mass distribution is essential, to avoid drawing an incorrect conclusion about the validity of the distribution. This is especially important in areas where there are extremely tight constraints, such as those from CMB anisotropies and evaporation, of particular interest for the LIGO and asteroid mass windows respectively. For cases involving fitting to accurate data or tight constraints, we advocate the use of the skew-lognormal or generalised critical collapse model, to ensure that the conclusions drawn are valid.

If the shape of the power spectrum peak deviates from that considered here, either by considering other symmetric peaks or non-symmetric peaks, the shape of the numerical mass distribution will naturally alter as well. We have shown that for the case of narrow peaks, the shape of the mass distribution is dominated by the effects of critical collapse, and becomes insensitive to the details of the power spectrum peak. Therefore, we do not expect the conclusions on the fitted models to change in this regime. Additionally, even a precisely measured mass distribution may only be able to confirm the critical scaling for PBHs, but not say anything about the underlying power spectrum peak. For much broader peaks, the detailed shape of the peak will be important, and may affect the results stated here. Nonetheless, we believe that in general, a three-parameter model will be required to capture the full shape of the mass distribution across a broad range of widths.

\section*{Acknowledgements}
AG is funded by a Royal Society Studentship by means of a Royal Society Enhancement Award. CB acknowledges support from the Science and Technology Facilities Council [grant number ST/T000473/1]. AH is supported by a Science and Technology Facilities Council Consolidated Grant.

\appendix
\renewcommand\thetable{A.\arabic{table}}
\renewcommand{\theHtable}{A.\arabic{table}}
\setcounter{table}{0}

\section{Reduced \texorpdfstring{$\chi^2$}{chi-squared} values}
\begin{table}[H]
\centering
\caption{$\chi^2_\nu$ values for different models and power spectrum widths.}
\begin{tabular}{c|ccccccc}
& \multicolumn{7}{c}{\textbf{Model}} \\
\textbf{Width} $\boldsymbol{\Delta}$ & Lognormal & Gaussian & Skew-normal & Skew-lognormal & CC1 & CC2 & CC3 \\ \hline
$\delta$ & $3.67\times10^{-4}$ & $4.15\times10^{-5}$ & $1.03\times10^{-5}$ & $8.55\times10^{-6}$ & $3.37\times10^{-4}$ & $9.88\times10^{-6}$ & $5.78\times10^{-7}$ \\

$0.1$ & $3.58\times10^{-4}$ & $4.56\times10^{-5}$ & $1.01\times10^{-5}$ & $8.32\times10^{-6}$ & $3.97\times10^{-4}$ & $1.16\times10^{-5}$ & $6.72\times10^{-7}$ \\

$0.3$ & $2.96\times10^{-4}$ & $8.76\times10^{-5}$ & $8.54\times10^{-6}$ & $6.61\times10^{-6}$ & $1.06\times10^{-3}$ & $2.93\times10^{-5}$ & $1.39\times10^{-6}$ \\

$0.5$ & $2.16\times10^{-4}$ & $1.91\times10^{-4}$ & $6.58\times10^{-6}$ & $4.32\times10^{-6}$ & $3.00\times10^{-3}$ & $7.11\times10^{-5}$ & $1.98\times10^{-6}$ \\

$1.0$ & $8.25\times10^{-5}$ & $7.15\times10^{-4}$ & $2.02\times10^{-5}$ & $9.33\times10^{-7}$ & $1.62\times10^{-2}$ & $2.56\times10^{-4}$ & $1.04\times10^{-6}$ \\

$2.0$ & $5.57\times10^{-6}$ & $2.41\times10^{-3}$ & $3.98\times10^{-4}$ & $8.90\times10^{-8}$ & $7.01\times10^{-2}$ & $7.09\times10^{-4}$ & $5.47\times10^{-8}$ \\

$5.0$ & $6.93\times10^{-5}$ & $5.51\times10^{-3}$ & $3.40\times10^{-3}$ & $2.90\times10^{-6}$ & $1.47\times10^{-1}$ & $1.74\times10^{-3}$ & $1.11\times10^{-4}$ \\ \hline
\end{tabular}
\label{tab:stat_values}
\end{table}

\bibliographystyle{JHEP-edit} 
\bibliography{Accurate_mass_distribution_model_paper}{}

\end{document}